\documentclass[10pt,conference,a4paper]{article}

\usepackage{authblk}
\usepackage{cite}
\usepackage{siunitx}
\usepackage{amsmath,amssymb,amsfonts}
\usepackage{algorithmic}
\usepackage{graphicx}
\usepackage{ltxgrid}
\usepackage{textcomp}
\usepackage{xcolor}
\usepackage{geometry}
\def\BibTeX{{\rm B\kern-.05em{\sc i\kern-.025em b}\kern-.08em
    T\kern-.1667em\lower.7ex\hbox{E}\kern-.125emX}}
\usepackage{flushend}

\date{} 

\begin{document}

\title{Expanding Scanning Frequency Range of \\Josephson Parametric Amplifier Axion Haloscope Readout with Schottky Diode Bias Circuit}

\author[1,2]{Minsu Ko}
\author[2]{Sergey V. Uchaikin}
\author[2]{Boris I. Ivanov}
\author[1,2]{JinMyeong Kim}
\author[2]{Seonjeong Oh}
\author[2]{Violeta Gkika}
\author[1,2]{Yannis K. Semertzidis}
\affil[1]{\small{Korea Advanced Institute of Science and Technology, Daejeon, South Korea}}
\affil[2]{Center for Axion and Precision Physics Research of Institute for Basic Science, Daejeon, South Korea}

\maketitle
\begin{abstract}
    The axion search experiments in the microwave frequency range require high sensitive detectors with intrinsic noise close to quantum noise limit. Josephson parametric amplifiers (JPAs) are the most valuable candidates for the role of the first stage amplifier in the measurement circuit of the microwave frequency range, as they are well-known in superconducting quantum circuits readout. To increase the frequency range, a challenging scientific task involves implementing an assembly with parallel connection of several single JPAs, which requires matching the complex RF circuit at microwaves and ensuring proper DC flux bias. In this publication, we present a new DC flux bias setup based on a Schottky diode circuit for a JPA assembly consisting of two JPAs. We provide a detailed characterization of the diodes at cryogenic temperatures lower than 4~\si{\kelvin}. Specifically, we selected two RF Schottky diodes with desirable characteristics for the DC flux bias setup, and our results demonstrate that the Schottky diode circuit is a promising method for achieving proper DC flux bias in JPA assemblies.
    \end{abstract}

\section{Introduction}
Axion search experiments are fundamental physical research dedicated to resolve the strong charge parity (CP) problem and describe the nature of dark matter\cite{peccei1977cp}.
The haloscope experiments, which use high-quality 3D cavities at high magnetic fields\cite{sikivie1983experimental, brubaker2017first}, are performed in the microwave frequency range\cite{Yi23} at the Center for Axion and Precision Physics Research (CAPP). 
The total system noise in the experiments is close to the quantum noise limit\cite{Caglar21, Yi23}.
This achieves the sensitivity of the Dine-Fischler-Srednicki-Zhitnitsky axion-to-microwave photon coupling model\cite{Yi23}. The key element of such a measurement chain is a JPA. 
The JPAs used for axion search experiments at CAPP, at frequencies around 1~GHz, have a tunable range from 30~\si{\mega\hertz} to 40~\si{\mega\hertz}, with a gain of more than 20~dB\cite{Caglar21, Yi23}.

\vspace{6pt}
The haloscope axion search experiments with a high magnetic field of 12~\si{\tesla} are performed using a dilution refrigerator inserted into the liquid He dewar. 
Such experiments are long-term, high-cost, and require scanning in a wide frequency range within a single cooldown of the dilution refrigerator (DR), with the experiment temperature less than 30~\si{\milli\kelvin}\cite{Yi23}.
This defines the requirements for the microwave detection chain to scan in a wide frequency range. 
In our particular case, introducing more DC lines into the dilution refrigerator with a 1~K pot cooled by liquid helium flow increases the overall heat load and reduced the efficiency of the mixture condensing.
A scheme with a connection of several JPAs with a single flux bias coil was designed, implemented, and applied in the axion search experiment performed at CAPP. 
Using this scheme, the scanning frequency range increased up to 120~\si{{\MHz}}~\cite{JinMyeong2023}. 

\vspace{6pt}
In order to implement a multiple JPA connection scheme, to simplify frequency adjustment and avoid interference between different JPA, it is often necessary to separate the flux bias for each JPA using different twisted pairs of wires. 
However, adding more wires to an experimental fridge can be challenging due to limited space and a limited number of fridge connectors.
To address this issue, we developed and implemented a new circuit design using two Schottky diodes.
This circuit enabled us to apply two dc flux biases independently to two JPAs using the same twisted pair of wires, by separating the flux biases using currents of different directions to bias one or the other JPA. 
The circuit incorporated a diode rectifier that operated at the cold stage of the fridge. 
We tested the diodes at both 300~\si{\kelvin} and 4~\si{\kelvin} to obtain their current-voltage characteristics (I--V curves). 

\section{Theory}
The I--V curve of diodes is governed by simplified Shockley diode equation\cite{shockley1949theory}:
\begin{equation}
I_D = I_S(e^{\frac{V_D}{nV_T}}-1),\label{eq}
\end{equation} 

\noindent here, $I_D$ and $I_S$ are the diode current and reverse-bias saturation current, respectively. 
$n$ is the ideality factor, $V_D$ is the voltage across the diode, and $V_T$ is the thermal voltage, which is given by $q_eV_T=k_BT$.
Where, the $q_e$ is the elementary charge, $k_B$ is the Boltzmann constant.
The equation describes the behavior in the forward-bias region and some part of the reverse-bias region. 
As shown in the Shockley equation, the IV behavior of a diode depends on temperature. 
As temperature increases, the threshold voltage decreases\cite{8854889} since the high temperature helps to excite the carriers and the probability of overcoming the barrier increases. 
On the other hand, the operation of each particular type of Schottky diodes at cryogenic temperatures necessitates a specialized investigation.

\vspace{6pt}
The JPAs used in our experiments are designed based on aluminum technology and are operated at temperatures below 50~\si{\milli\kelvin}, where the cooling power of a dilution fridge is insufficient to handle the power dissipation of the diodes. 
Therefore, the diode circuit should be mounted on the still, 1~\si{\kelvin}, or 4~\si{\kelvin} stage of the fridge. 
The properties of Schottky diodes at cryogenic temperatures can change dramatically, necessitating a preliminary experimental study. 
It has been demonstrated that the Schottky barrier allows the diode to respond quickly to current switching and operate correctly at low temperatures~\cite{zheng2021dynamic}. 
Previous researches have tested Schottky diodes at temperatures and characterized their properties including I--V behavior and hole diffusion length near 100~\si{\kelvin}~\cite{I2, hardikar1999anomalous, vittone2007temperature}.
One of the results indicated a satisfactory level of concordance between the simulations based on theories and measurements, particularly at least in the vicinity of 100~\si{\kelvin}~\cite{I1}.
More recently, the characteristics of Schottky diodes have been examined in extreme environments, including strong magnetic field~\cite{magnetic} and cryogenic temperatures~\cite{shim2021cmos}.
One of them revealed the expected behavior down to 77~\si{\kelvin} with an ideality factor close to one.
It is shown that the magneto-resistance of the diode slowly drops with increasing of the magnetic field up to 6~\si{\tesla} at both 4.2~\si{\kelvin} and 77~\si{\kelvin}.
The objective of this experiment is to design a circuit utilizing RF Schottky diodes that can regulate the current direction in the DC flux bias circuit of JPAs operating at temperatures of 4~\si{\kelvin} and below.

\section{Experimental Setup}
\begin{figure}
\centering
\includegraphics[width=0.7\textwidth]{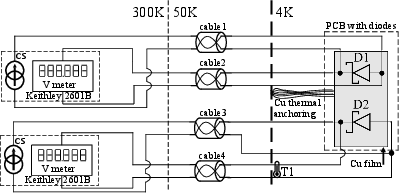}
\caption{A simplified measurement setup based on the dry closed cycle cryocooler. D1-D2: Schottky diodes placed on the PCB and thermally anchored to 4~K stage; cables from 1 to 4: twisted pair brass cables in CuNi shiled; Cu film: a copper sheet soldered to the PCB for better thermal anchoring of the diodes; T1: LakeShore Cernox temperature sensor; CS: current source and V meter: voltage meter, both are integrated into the Keithley 2601B sourcemeter.}
\label{fig:1}
\end{figure}

The SMS7630-040LF radio frequency (RF) Schottky diodes were experimentally studied at 300~\si{\kelvin} and about 3~\si{\kelvin}.
The Keithley 2601B precision source measure unit and the Lakeshore~325 cryogenic temperature controller were used in order to measure I--V curves and LakeShore Cernox temperature sensor factory calibrated down to 1~\si{\kelvin}. 
The measurements were performed in the dry closed cycle cryocooler based on a two-stage pulse tube and Cryomech compressor.
The experimental setup is shown in Figure~\ref{fig:1}.
In order to mount the diodes to the cryostat and perform the measurements, the samples were soldered to the PCB and placed at the 4~\si{\kelvin} stage of the cryocooler. 
The additional thermal anchoring for the samples was provided by copper braided wire, shown as Cu thermal anchoring in Figure~\ref{fig:1}.
In order to improve the connection between the diodes and the copper braided wire, we placed additional copper sheet.
This copper sheet is shown as Cu film in Figure~\ref{fig:1}. 
The DC connection starts from the room temperature Fischer connector at 300~\si{\kelvin}, which goes to the 24pin connector at 50~\si{\kelvin} stage.
From the 50~\si{\kelvin} stage to the 4~\si{\kelvin} stage, the cabling was done using twisted pair brass cables in CuNi shields, shown as cable1 to cable4 in Figure~\ref{fig:1}.
The shields of the cables were thermally anchored at each temperature stage.

\section{Experimental Study}
\begin{figure}[t]
   \begin{minipage}{0.48\textwidth}
     \centering
     \includegraphics[width=\linewidth]{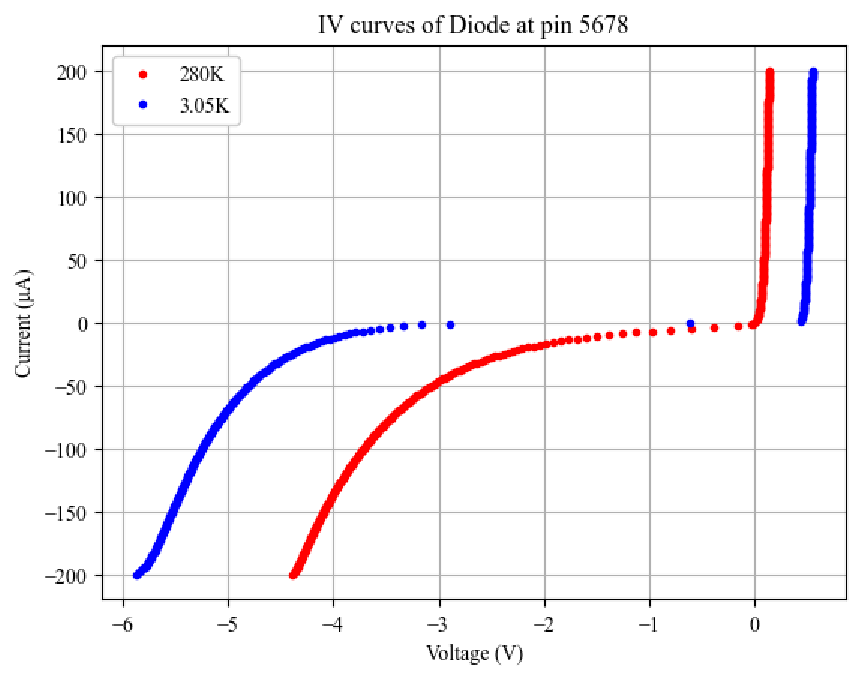}
     \caption{The I--V curves of the diode at 280~\si{\kelvin} and 3.05~\si{\kelvin} in a current range from -200~\si{\micro\ampere} to 200~\si{\micro\ampere}.}\label{fig:2}
   \end{minipage}\hfill
   \begin{minipage}{0.48\textwidth}
     \centering
     \includegraphics[width=\linewidth]{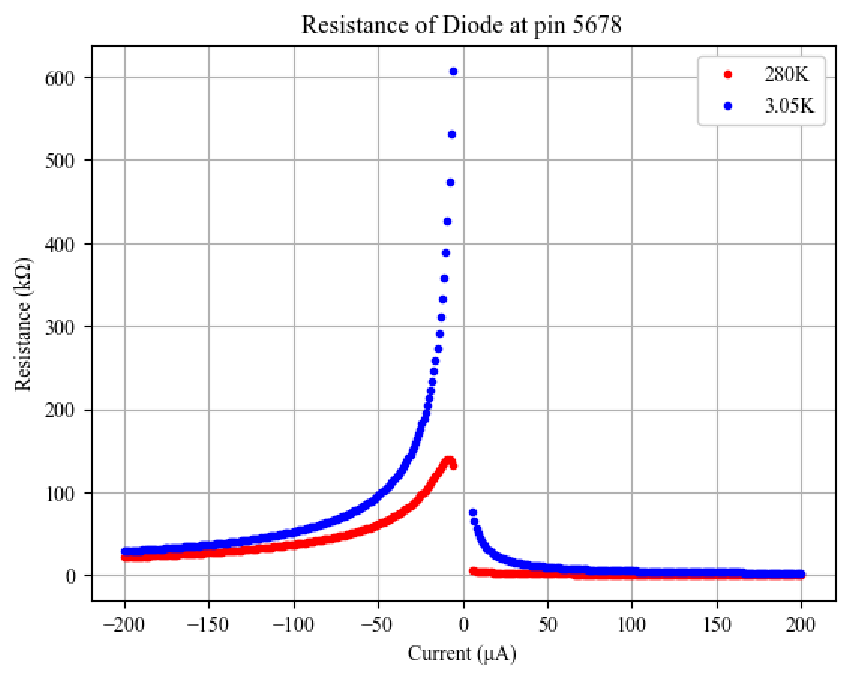}
     \caption{The resistance of the diode at 280~\si{\kelvin} and 3.05~\si{\kelvin} in a current range from -200~\si{\micro\ampere} to 200~\si{\micro\ampere}.}\label{fig:3}
   \end{minipage}
\end{figure}

The measurement procedure involved sweeping the DC current over the diode and measuring the resulting voltage. 
The 4-point measurements were performed on two diodes at a stable cryogenic temperature of 3.05~\si{\kelvin}. 
The diode measurements were conducted over a range of -200~\si{\micro\ampere} to 200~\si{\micro\ampere}, which corresponds to the bias current range of the JPAs used in CAPP axion search experiments. 
The measured I--V curves of the first Schottky diode at 300~\si{\kelvin} and 3.05~\si{\kelvin} are shown in Figure~\ref{fig:2}. 
It was observed that the Schottky diode exhibited higher forward voltage drop and breakdown voltage at cryogenic temperatures than at room temperature. 
The resistance of the diode at two different temperatures was obtained and is shown in Figure~\ref{fig:3}.
\newpage

\begin{figure}[htb!]
\centering
\includegraphics[width=0.4\textwidth]{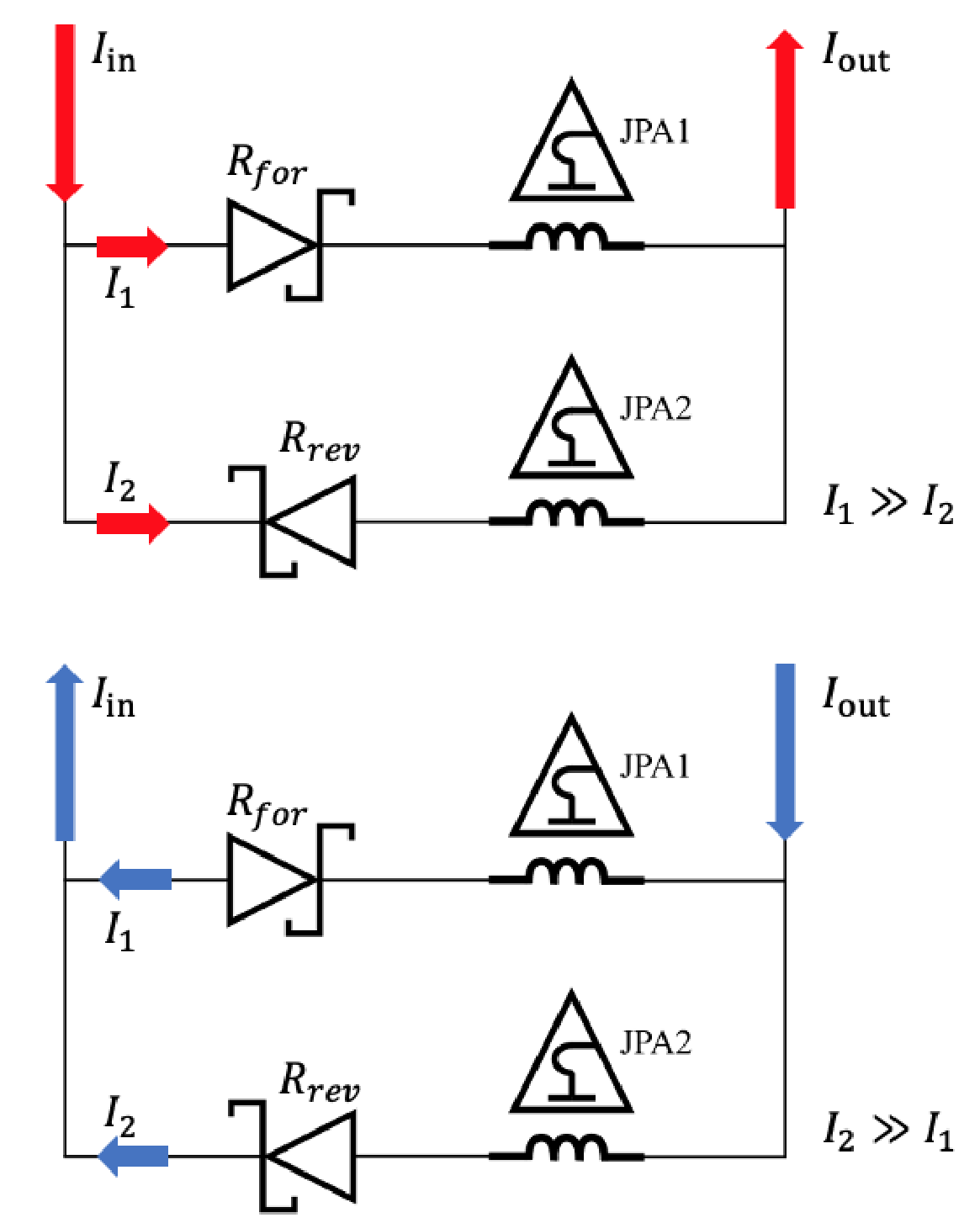}
\caption{The simplified schematic of the two JPAs bias circuit with two diodes connected in parallel. $I_{\text{in}}$: the total current applied to the circuit, $I_1$, $I_2$: splitted currents, $R_{for}$, $R_{rev}$: resistances for forward and reversed bias. The figure depicts the currents flowing in both directions.}
\label{fig:4}
\end{figure}
Based on our measurements, the tested diodes are capable of serving as rectifiers throughout the entire temperature range between 3 and 300~\si{\kelvin}.
To quantitatively compare the rectification performance at room temperature and 3.05~\si{\kelvin}, we estimated the parallel rectification factor $\rho_p$. 
The simplified schematic with two diodes connected in parallel and two JPAs is shown in Figure~\ref{fig:4}.
The total current $I_{\text{in}}$ is divided into $I_1$ and $I_2$ to provide an equivalent voltage to both diodes, where the resistance of each diode is denoted as $R_{\text{for}}$ and $R_{\text{rev}}$, respectively. 
The dependence of $R_{\text{for}}$ and $R_{\text{rev}}$ on $I_{1}$ and $I_{2}$ is shown in Figure~\ref{fig:3}. 
The rectification factor of the diodes is represented by the parameter $\rho_p$, which would ideally equal 1, indicating full isolation with no current flowing over the reversed diode.
\begin{equation}
\rho_{p} = \frac{I_{1}}{I_{\text{in}}}.
\end{equation}
The estimation of $\rho_p$ is based on the assumption that the voltages across the two diodes are equivalent and that the circuit follows Kirchhoff's law, which is a valid assumption since both JPA flux bias coils are superconducting.
The system of equation follows:
\begin{equation}
\begin{cases}
I_{1}R_{for}(I_{1}) = I_{2}R_{rev}(I_{2}), \\
I_{\text{in}} = I_{1}+I_{2}. \\
\end{cases}
\end{equation}
To estimate of the $\rho_p$ we need to obtain one of the currents $I_{1}$ or $I_{2}$.
For the applied bias current $I_{\text{in}}=200~\si{\micro\ampere}$, the resistance values from Figure~\ref{fig:3} and solving the system of equations we obtained $I_{1} = 197.43~\si{\micro\ampere}$ at 280~\si{\kelvin} and $I_{1} = 198.54~\si{\micro\ampere}$ at 3.05~\si{\kelvin}.
This yields $\rho_p = 0.987$ at 280~\si{\kelvin} and $\rho_p = 0.993$ at 3.05~\si{\kelvin}.
It means that 99.3\% of the total current flows through the forward bias diode, and only 0.7\% of the total current is the leakage at 3.05~\si{\kelvin}, while 1.3\% of the total current is leakage at 280~\si{\kelvin}.
It shows that the RF Schottky diode has better performance at 3~\si{\kelvin} than at room temperature.

\begin{figure}
\centering
\includegraphics[width=0.5\textwidth]{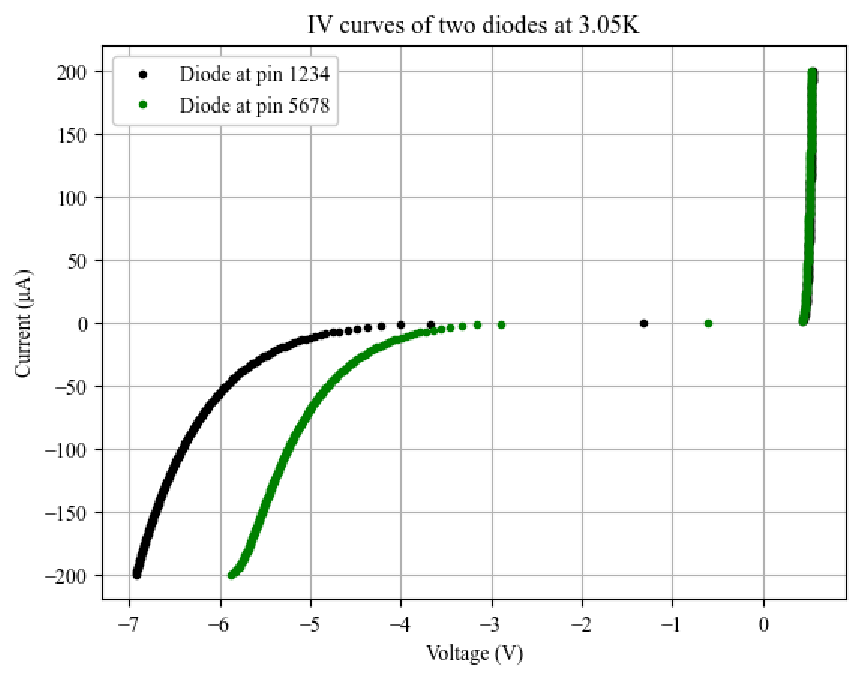}
\caption{I--V curves of two diodes measured during one experiment at 3.05~\si{\kelvin}.}
\label{fig:5}
\end{figure}

\begin{figure}
\centering
\includegraphics[width=0.5\textwidth]{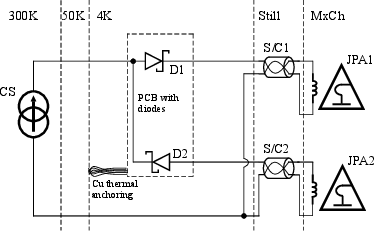}
\caption{Simplified schematic of the JPA DC flux bias line, D1-D2: Schottky diodes placed on the PCB and thermally anchored to 4~K stage; s/c1, s/c2: twisted pair superconducting NbTi cables in CuNi shield, MxCh: mixing-chamber plate, and CP: cold plate. The s/c1, s/c2 are thermally anchored at MxCh, Still and 4~\si{\kelvin} plate}
\label{fig:6}
\end{figure}

\vspace{6pt}
We performed measurements on several diodes for our JPA circuit, and the I--V curves of two of these diodes measured at 3.05~\si{\kelvin} are shown in Figure\ref{fig:5}. 
Both diodes exhibit desirable characteristics in forward bias, with no significant difference between their I--V curve behaviors. 
However, their breakdown voltages in reverse bias differ by about 1~\si{\volt}, as one can see in Figure~\ref{fig:5}. 
The slight difference in breakdown voltage between the two characterized diodes does not affect the operation of our JPA flux bias circuit since we use only the positive direction of the DC current and the reverse current leakage is around 1 \% for 200~\si{\micro\ampere}.
Therefore, we proceeded with implementing the characterized diodes in the JPA flux bias circuit, and a simplified schematic of the circuit applied for the JPA bias line is shown in Figure~\ref{fig:6}.
\newpage

\section{Conclusion}
The RF Schottky diodes were thoroughly characterized at the temperature of 3.05~\si{\kelvin} using the closed cycle cryocooler.
The parallel rectification factor was defined and used to quantitatively compare the RF Schottky diodes at room and cryogenic temperatures, which showed that the diodes are capable of rectification down to 3~\si{\kelvin}. 
These characterized diodes are currently being used in the JPA flux bias circuit for the CAPP-MAX axion search experiment, which is based on a 12~\si{\tesla} magnet and carried out at the Center for Axion and Precision Physics Research of Institute for Basic Science. 

\vspace{6pt}
The RF diodes circuit shown in the current publication allowed us to implement a multiple JPA connection scheme and simplified frequency adjustment, which is necessary for axion search scanning in a wide frequency range. 
This scheme allows to avoid interference between different JPAs by means of the flux bias separation and to reduce the number of DC lines introduced to the dilution refrigerator.
This causes the reducing of the total heat load to the refrigerator. 
The implemented circuit enabled us to apply two DC flux biases to two JPAs independently using the same twisted pair cabling.
In particular, it was done by separating the flux biases using different directions of DC currents for one or the other JPA.
As a further application we consider to install the proposed diode circuit to the axion search readout based on the 6 JPAs.
\section*{Acknowledgement}
This work is supported by the Institute for Basic Science IBS-R017-D1.

\bibliographystyle{plain}
\bibliography{REFERENCES}

\end{document}